\documentclass[12pt,preprint]{aastex}
\usepackage{emulateapj5}

\usepackage{amssymb}

\slugcomment{Submitted in the ApJ on April 15, 2003}
\shorttitle{A Chandra/SIRTF Starburst high-z diagnostic}
\shortauthors{Weedman et al.}

\begin{document}


\title{Comparing Chandra and SIRTF Observations for Obscured Starbursts and AGN  at High Redshift}

\author{D. Weedman, V. Charmandaris\altaffilmark{1}}
\affil{Astronomy Department, Cornell University, Ithaca, NY 14853}
\email{dweedman@isc.astro.cornell.edu,vassilis@astro.cornell.edu}

\author{A. Zezas}
\affil{Harvard-Smithsonian Center for Astrophysics, 60 Garden Street, Cambridge, MA 02138}
\email{azezas@head-cfa.harvard.edu}

\altaffiltext{1}{Chercheur Associ\'e, Observatoire de Paris, LERMA, 61 Av. de l'Observatoire, F-75014 Paris, France}

%\newpage

\begin{abstract}

Tracking the star formation rate to high redshifts requires knowledge
of the contribution from both optically visible and obscured sources.  The dusty, optically-obscured galaxies can be
located by X-ray and infrared surveys. To establish criteria for selecting such sources based only on X-ray and infrared surveys, we determine the ratio of
infrared to X-ray brightness that would be observed by SIRTF and
Chandra for objects with the same spectral shapes as nearby starbursts
if seen at high redshift.  The parameter IR/X is defined as IR/X = (flux density observed in SIRTF
MIPS 24\,$\mu$m filter in mJy)/(total flux observed within
0.5--2.0\,keV in units of 10$^{-16}$ ergs s$^{-1}$ cm$^{-2}$).  Based
on observations of NGC\,4038/39 (``The Antennae''), NGC\,3690+IC\,694
(Arp\,299 or Mkn\,171), M\,82, and Arp\,220, nine 
starburst regions are compared using mid-infrared spectra taken by the
Infrared Space Observatory (ISO) and X-ray spectra obtained with
Chandra . The IR/X are determined as they would appear for $1<z<3$.
The mean IR/X over this redshift range is 1.3 and is not a significant
function of redshift or luminosity, indicating that SIRTF surveys
reaching 0.4\,mJy at 24\,$\mu$m should detect the same starbursts as
deep CXO surveys detect at a flux of
0.3$\times$10$^{-16}$\,ergs\,s$^{-1}$\,cm$^{-2}$.  The lower bound of
IR/X for starbursts is about 0.2, suggesting that objects with IR/X smaller than this have an
AGN X-ray component in addition to the starburst. Values of IR/X for
the obscured AGN within NGC\,1068, the Circinus galaxy, and NGC\,6240
are also determined for comparison although interpretation is complicated
by the circumnuclear starbursts in these galaxies. Any sources found in surveys having IR/X$>$4 would not match any of the objects considered.  

\end{abstract}


\keywords{dust, extinction ---
	 galaxies: high-redshift --
	infrared: galaxies ---
	X-rays: galaxies ---
        galaxies: AGN ---
	galaxies: starburst}

\section{Introduction}

Understanding the formation and evolution of galaxies requires
measuring the star formation rate as a function of redshift in the
universe.  Currently available data indicate that the star formation
rate was much greater at z$\sim$1 than at the present, but whether
this rate continues to increase or stabilizes for z$>$1 is essentially
unknown \citep{Chary01}. An equally crucial issue is understanding the
development and evolution of active galactic nuclei (AGN) and how this
relates to star and galaxy formation.  An important challenge for
these questions is to determine the fraction of sources that are so
obscured by dust that they have not been accounted for in optical
surveys.  Such objects are probably the dominant source of observed
background emission at infrared and submillimeter wavelengths and
seemingly comprise the majority of the luminosity density in the
universe \citep[e.g.][]{Hauser98,Barger00}.  Even when optically obscured, dusty sources can be selectively located by X-ray surveys
and by infrared surveys.  In X-rays, radiation from the primary source
can be seen directly as long as the HI column density of the obscured
region is less than $\sim$10$^{24}$ cm$^{-2}$ or can be seen in X-rays
scattered outside the obscuring region.  In the infrared, dust is seen
via direct emission of infrared luminosity; this luminosity arises from reradiation by the dust of energy initially
emitted as  X-ray, ultraviolet  and visible photons. 

It is established already from deep surveys with the Chandra X-ray
Observatory (CXO) that a substantial population of objects exist which
are too faint for optical spectroscopic redshift determination.  For
example, about 10\% of the sources in the Chandra Deep Field-North
survey (CDF-N) are optically blank, with R$>$26.5\,mag
\citep{Hornschemeier01}.  These may be optically faint because of
obscuration by dust, which becomes even more significant at high
redshift when the observed optical wavelength corresponds to
ultraviolet wavelengths in the source rest frame.  The CDF-N has also
shown that CXO sources are often the same as infrared sources found in
deep surveys with the Infrared Space Observatory (ISO)
\citep{Alexander02}.  In the overlapping survey area (about
22\,arcmin$^2$, or 5\% of the CDF-N area), there are 41 ISO sources and 49 CXO sources, and about one-third of these are
seen by both.  This is an important indication that both X-ray and
infrared surveys are crucial in identifying dust
enshrouded distant galaxies.

The next opportunity for major progress in infrared surveys will be
with the Space Infrared Telescope Facility
(SIRTF)\footnote{Information on SIRTF is at
http://sirtf.caltech.edu/SSC/}. It is the intention with SIRTF to undertake surveys that can detect infrared sources
over areas of many
square degrees, and all of the CXO deep fields are planned to be covered by SIRTF.  Because CXO and SIRTF will continue simultaneous operation for
many years, it is probable that numerous regions of the extragalactic
sky will be imaged both in X-rays and infrared.  SIRTF will readily
extend to higher redshifts the infrared census previously determined
by ISO.  SIRTF surveys will be especially sensitive to dusty
starbursts because of the strong suite of spectral features between 6
and 13\,$\mu$m rest wavelength arising from molecular emission by
polycyclic aromatic hydrocarbons (PAH), features which are very strong
in starbursts (e.g.  Figure 2).  These emission features are so strong
that an object can appear many times brighter in images at wavelengths
centered on this emission than in images at continuum wavelengths.  For example, the
strong 7.7\,$\mu$m feature made the extensive surveys by ISOCAM at
15\,$\mu$m particularly sensitive to starburst galaxies at z$\sim$1
\citep{Elbaz02}.  The analogous SIRTF surveys will be at
24\,$\mu$m, using the Multiband Imaging Photometer (MIPS), and will
readily reach survey limits of 0.4\,mJy in large area, shallow surveys
\citep{Dole03}.  This will allow the extension to z$\sim$3 of the
luminosity function for dusty starbursts tracked to z$\sim$1.2 by
ISO.  For obscured starbursts too faint for optical redshift
determination, the Infrared Spectrograph on SIRTF will be able to
determine redshifts to z$\sim$4 based on the strong PAH features.

Both AGN and starbursts produce copious X-ray and infrared emission,
but for dramatically different reasons.  Hard X-rays arise from hot
accretion disks near massive black holes in AGN, but in starburst
galaxies they originate from accretion disks near compact stellar
remnants; the summed contribution of these individual sources makes starbursts a significant source of hard X-rays. Soft X-rays arise from hot winds flowing from accretion
disks in AGN but also arise from young stars, stellar winds, and supernova
remnants in starbursts.  Dust reradiation comes from dust grains and
molecules which are either heated or photoexcited by the radiation
from accretion disks in AGN, or from hot stars in starbursts.  While
it is often assumed that most extragalactic X-ray sources are AGN, it
is known that ultraluminous starburst galaxies can be significant
X-ray sources even at high redshift \citep[e.g.][]{Moran99,Zezas98,Hornschemeier02},
and the majority of faint sources with both CXO and ISO detections
have been interpreted as starburst galaxies rather than AGN
\citep{Alexander02}.  It is important to determine how the comparison
of SIRTF and CXO surveys can locate those luminous, dusty sources
powered primarily by starbursts and distinguish them from sources
powered by AGN.

The redshift interval 1$<$z$<$3 is particularly crucial to explore for
several reasons.  It is known from optical surveys that this is the
interval in which the optically discoverable quasar luminosity
function peaks \citep{Fan01}.  It might be expected, therefore, that
significant numbers of X-ray selected AGN would also be found at these redshifts, and the X-ray surveys should detect any significant
population of obscured sources. This is also the interval in which
modeling of infrared source counts implies that ultraluminous infrared
galaxies powered by starbursts reach their peak \citep{Xu03}.
Determining luminosity functions out to z$\sim$3 that include obscured
sources is the only way to determine when both star formation and AGN
activity peaked in the universe.  There are adequate reasons to expect
that SIRTF will detect many sources at z$>$1.  Various infrared and
submillimeter observations demonstrate that ultraluminous infrared
galaxies (ULIRGs) exist with bolometric luminosities about
10$^{13}$\,L$_{\sun}$.  Source counts and evolution deduced from
background constraints require that such sources are hundreds of times
more common per unit volume at z$>$2 than in the local universe
\citep[e.g.][]{Lagache03}. Taking the spectral energy distribution of
a 10$^{13}$\,L$_{\sun}$ galaxy from \citet{Chary01} along with their
cosmological assumptions, the observed flux density at 24\,$\mu$m for
z=2 would be 2.7\,mJy (enhanced by a strong PAH feature); at z=3, this
flux density would be 0.4\,mJy.  Because SIRTF surveys will readily
detect sources to these flux limits, SIRTF will have the potential to
produce a luminosity function for obscured sources within a redshift
range that is crucial for tracing ULIRGs at redshifts around the star-formation peak. It is important to determine how these ULIRGs might relate
to faint, high redshift sources revealed in X-ray surveys and to learn
if there is a selection criterion to distinguish between starburst and
AGN as the dominant power source. Using comparisons of infrared and X-ray observations to consider this
question is the objective of the current paper.

A necessary step toward using comparisons of infrared and X-ray measurements for
studying the high redshift universe is to determine the relative
``K-corrections'' in the different wavelength regimes, because
rest-frame emissions are found at significantly different observed
wavelengths at high redshift.  For example, by z=2, the CXO ``soft
band'' (0.5--2.0\,keV)
actually samples the rest-frame energies of 1.5--6\,keV. The deepest dust-sensitive surveys by SIRTF will be at an
observing wavelength of 24\,$\mu$m, corresponding to rest frame of
8\,$\mu$m for z=2.  To determine criteria for comparing CXO and SIRTF
sources for the purpose of identifying candidate high redshift
starburst galaxies, and for comparing starbursts with AGN, we need to
know the characteristics of pure starbursts in the X-ray and infrared
spectral regimes over broad enough wavelength ranges to determine
relative K-corrections.  To accomplish this, we consider in this paper
a detailed comparison of the X-ray (from CXO) and infrared properties
(from ISO) in the best examples of nearby, pure starbursts which have
observations extending to the higher X-ray energies and shorter infrared wavelengths
where such objects would be seen in the observer's frame at high
redshifts.  Our specific objective is to determine the range of
infrared/X-ray ratios which should be observed for pure starbursts at
various redshifts up to z=3, when observed in the CXO soft band, the
most sensitive CXO band, and the SIRTF 24\,$\mu$m MIPS band.  By using
known examples of local starbursts, these ratios can be determined
empirically.  We also use a more limited sample of obscured AGN as an
initial step toward comparing them with starbursts.

\section{Choice of objects for study}

The ``pure starburst'' objects used for these infrared to X-ray
comparisons are luminous galaxies in both X-ray and infrared for which
careful multiwavelength studies have shown that they only harbor
luminous starbursts with no evidence for AGN activity.  For
our comparisons, we also require that adequate archival ISO and CXO
data exist in order to determine what the infrared/X-ray ratio would
be if observed at various redshifts.  Our approach is strictly
empirical, so we desire to include objects with a variety of
morphologies and luminosities, including nuclear starbursts, entire
interacting systems, and individual starbursts within galaxies.  For
the distant objects to be sought with SIRTF surveys, the spatial resolution will not be
adequate to distinguish among these different types of starbursts based on morphology.

Archival data are now available from CXO and ISO for a number of local
starburst galaxies for which spectral data can be extracted from
spatial areas identical in both X-ray and infrared. 
The objects used in the present paper are the interacting system
NGC\,3690+IC\,694 (also Markarian 171 or Arp\,299), the interacting
system NGC\,4038+NGC\,4039 (also the ``Antennae''), the prototypical
starburst galaxy M\,82, and Arp\,220, the prototypical ULIRG.  Taking
this collection of objects, we will be able to describe the
infrared/X-ray ratios of 9 different starburst regions with sufficient
spectral coverage to describe how they would appear if they were
located at 1$<$z$<$3.  The regions used cover a luminosity range
approaching 100 and thus will allow investigation of whether the ratio
depends on luminosity.  ISOCAM 15\,$\mu$m images of the sources
showing the spatial regions compared in infrared and X-ray are shown in Figure
1. The appearance of these regions in X-rays depends on the energy, and representative images illustrating this are cited below in discussions of individual objects.  

1. NGC\,3690 + IC\,694: These combined objects comprise the system Arp\,299 or Markarian 171 which has long
been interpreted as a luminous starburst galaxy \citep{Gehrz83}.
Careful studies utilizing multiwavelength capabilities including HST
have verified that virtually all of the luminosity at all wavelengths
arises from multiple starbursts and their consequences
\citep{Zezas98,Alonso00,Charmandaris02}.  ISO observations of this system have
sufficient spatial resolution to isolate numerous starbursts for
potential comparison with CXO observations
\citep{Gallais98}. An X-ray image showing the dramatic difference in appearance between soft band and hard band X-rays is in \citet{Zezas03}. We analyze three separate regions for this system: the components NGC\,3690 and IC\,694, and the total system. Even though the total system includes the other two regions, the system luminosity also includes extended emission which is outside of the smaller regions, so the extended total system is considered as a separate starburst region. Although
recent observations with Beppo Sax at very high energy X-rays
($>$10\,keV) reveal evidence for an AGN that may account for a few
percent of the bolometric luminosity \citep{Dellaceca02}, this AGN is very weak in the CXO band and does not affect the measured spectrum
and flux of NGC\,3690, where it is located (Zezas et al. 2003).  The
regions defined for our infrared and X-ray measures are illustrated in
Figure 1a. The distance adopted is 42\,Mpc.

2. NGC\,4038 + 4039: These objects comprise the ``Antennae'' system of interacting galaxies, which
provides another dramatic illustration of how the X-ray morphology of a
starburst system changes with energy \citep{Fabbiano01,Fabbiano03}.
The soft band (in the rest frame) reveals extended, diffuse emission,
whereas hard bands show individual sources with little diffuse
emission.  Optical, infrared, and millimeter observations thoroughly
demonstrate the starburst nature of all components of the system
\citep{Vigroux96,Mirabel98,Wilson00}, and there is no evidence for AGN activity anywhere within the system.  To accommodate empirically the
variety of morphologies, we examine 4 regions: the localized
``Knot A'', the nuclei of NGC\,4038 and 4039, and the entire system. As in Arp 299, the additional extended emission outside of the localized regions makes it appropriate to consider the entire system as a separate starburst region. The
regions defined for our infrared and X-ray measures are illustrated in
Figure 1b.  The distance adopted is 20 Mpc.

3. Arp\,220: This luminous infrared source is often described as the
nearest ULIRG, but detailed multiwavelength studies have found no
evidence for an AGN \citep[see][]{Smith98,Iwasawa01}, so it is considered as the
most luminous local starburst
\citep[e.g.][]{Sturm96,Charmandaris99,Clements02}. The luminosity is
concentrated in the nucleus, so we compare X-ray and infrared results
only for the nuclear region, illustrated in Figure 1c.  The distance
adopted is 76\,Mpc.

4. M\,82: This nearby galaxy has long been considered the prototype
for an extended starburst \citep{Rieke80}.  The interpretation of the
ISO mid-IR data used is presented by \citet{Forster03}.  CXO results
are intriguing but problematical for our current analysis, because of
the presence of a luminous, highly variable, hard source
\citep{Kaaret01,Matsumoto01}.  When in a high state, this dominates
the X-ray luminosity of the galaxy.  For our present analysis, we did
not include this source.  As a result, the X-ray fluxes which enter
our calculations for M\,82 in Table 1 could sometimes be a factor of
two higher than listed. As discussed further in section 4, below, M82 would be more similar to other starbursts in IR/X if the high-state flux were included.  Except for this single source, we compare
X-ray and infrared results for the entire galaxy illustrated in Figure
1b. Although this region does not include the galactic superwind seen in soft
X-rays \citep[e.g.][]{Watson84,Strickland97}, our redshifted
bands correspond to X-ray emission above 1 keV so the X-ray component seen in these bands is associated with the
optical body of the galaxy covered by our aperture.  The distance adopted
is 3.3\,Mpc.

\section{Analysis of infrared and X-ray data}

Because our approach is empirical, we undertake an analysis to produce
results directly applicable to observed measurements from CXO or
SIRTF, as would be derived from wide field imaging surveys.  For CXO,
results for the most sensitive observations are usually given as total
flux in the soft band from 0.5\,keV to 2\,keV.  For SIRTF, the most sensitive results for dust reemission from obscured regions 
will be flux densities in the 24\,$\mu$m MIPS band. The ratio of infrared
to X-ray brightness which we define in this paper to compare SIRTF and Chandra measurements is the parameter IR/X, defined as IR/X
= (infrared flux density measured with MIPS 24\,$\mu$m in
mJy)/(total X-ray flux observed between 0.5 and 2.0\,keV in
units of 10$^{-16}$ ergs s$^{-1}$ cm$^{-2}$).  In order to determine IR/X as would be observed at
different redshifts, the relevant bandpasses have to be transformed to
the rest frame spectra of the sources used.  The objects which we
evaluate have X-ray and infrared spectra extending over sufficient
wavelengths in the rest frame that IR/X can be determined for
redshifts 1$<$z$<$3.

The infrared spectra have various strong features which are comparable
in spectral width to the MIPS 24\,$\mu$m filter bandpass.  Because
observed results are given as a flux density at a single wavelength,
it is necessary to determine the flux density at the corresponding
rest-frame effective wavelength of the filter.  Spectra which make this possible are readily
available from the ISO archive. The infrared observations used in this
paper were obtained with ISOCAM \citep{Cesarsky96}, a 32$\times$32
pixel array on board the ISO satellite, as part of the CAMACTIV GTO
Program (principal investigator, F.  Mirabel).  Spectral maps of a
region around each target were created using the Continuously Variable
Filter (CVF) covering the full wavelength range 5.1--16.3\,$\mu$m with
a spectral resolution of $\sim$40.  For NGC\,4038/39, Arp\,220, and
NGC\,3690/IC\,694, the total field was $48''\times48''$ with 1.5$''$
pixels.  For M\,82, field size and pixel size were 92$''$ and 3$''$, respectively.
The spatial regions measured are shown in Figure 1.  In all cases, the
FWHM of the PSF varied between 4$''$ and 5$''$.  Data were analyzed
with the CAM Interactive Analysis software (CIA\footnote{CIA is a
joint development by the ESA astrophysics division and the ISOCAM
consortium}).  Since one of us (VC) was involved in the analysis and
publication of the ISO results of all galaxies included in our sample,
the reduced ISO data and analysis techniques we used were the same as
those of \citet{Vigroux96} and \citet{Mirabel98} for NGC\,4038/39,
\citet{Charmandaris99} for Arp\,220, \citet{Gallais98} for
NGC\,3690/IC\,694, and \citet{Forster03} for M\,82.  Spectra of all the sources
are shown in Figure 2.

The parameter IR/X is measured in the observer's frame so is given by IR/X = $f_{\nu}$(1+z)/$f_x$.  Here, $f_{\nu}$ is the flux density in the rest frame at the effective wavelength, $\lambda_{eff}$, observed with the SIRTF MIPS 24\,$\mu$m filter for the z of
interest. The $f_x$ is the total X-ray flux
within the rest-frame energy range of 0.5(1+z)\,keV to 2.0(1+z)\,keV.  Using the ISOCAM CVF spectra, the 24\,$\mu$m filter
response\footnote{SIRTF Observer's Manual;
http://sirtf.caltech.edu/SSC/obs/ } with all wavelengths divided by
(1+z) is folded into the spectrum to determine $\lambda_{eff}$ for various redshifts.
The source $f_{\nu}$ is then measured as the
flux density at this $\lambda_{eff}$ in the ISOCAM spectra. 

To determine $f_x$, each evaluation requires a different
energy cut for the CXO data, defining bands in the galaxy rest frames
which would correspond to 0.5--2.0\,keV in the observer's frame for
the different redshifts.  For example, at z=2, the observed
0.5--2.0\,keV soft band corresponds to an energy range of
1.5--6.0\,keV in the rest frame.  The $f_x$ were measured by fitting
spectra extracted from the CXO Advanced CCD Imaging Spectrometer
(ACIS) observations of the sources, obtained from the CXO archive, and
then calculating the total fluxes in the necessary bands from these
spectra.  The CXO data were analyzed following standard procedures
\citep[e.g.][]{Zezas02,Zezas03} using the CIAO v2.3 data analysis tool
suite. After screening for time intervals of high background, X-ray
spectra were extracted within the same spatial regions as for the ISO
spectra.  Because these regions are large compared to the scale over which ACIS response varies, we created response files for smaller segments where the response is constant and subsequently combined these to create the final response matrix and ancillary response files for a given region.    Spectral
fitting was performed with the XSPEC v11.1 package. It is desired to
 have only observed fluxes as they emerge from the sources, so no
 corrections were applied for internal absorption within the sources, although we
did correct for
the Galactic HI column density along the line of sight to each galaxy \citep{Stark92}. These Galactic corrections are negligible within the energy bands used.

Table 1 summarizes the infrared and X-ray observational results for $\lambda_{eff}$, $f_{\nu}$, and $f_x$.  The resulting 
IR/X are those which would be observed at the different redshifts listed for
spectra having shapes in the infrared and X-ray like those of the various starburst sources
utilized.  These results are also displayed in Figure 3.

\section{Comparison of SIRTF and CXO surveys}

The results in Table 1 and Figure 3 show the range of IR/X values as a
function of redshift for the different starbursts which are analyzed.
Table 2 summarizes the mean IR/X over all redshifts for the different
sources along with the source luminosities in the 2--8\,keV band
(arbitrarily chosen for luminosity comparison since it represents the
broadest bandpass utilized). Many different physical effects
can influence IR/X in different starbursts.  These effects
include what fraction of total radiated luminosity is absorbed by
dust, what fraction of bolometric luminosity emerges at the rest
wavelength observed in the infrared, the nature of the X-ray sources
compared to the sources powering the infrared, the shape of the
intrinsic X-ray spectrum, and how much the X-rays are absorbed. We do not attempt to pursue the implications of any of
these results as might regard the physical nature of starbursts.  Our
objective is strictly to present empirical criteria as a baseline for
future analysis of other sources.

The most important empirical results are: 
(a) The IR/X is not a significant function of redshift, except for
redshift of z=2, where the strong 7.7\,$\mu$m PAH feature produces
significant additional flux density within the 24\,$\mu$m filter band.
(b) The IR/X is also not a function of luminosity; for example,
NGC\,4039 and Arp\,299 starbursts have very similar values of IR/X
even though their X-ray luminosities differ by a factor of over 60.
(c) The IR/X ratio for starbursts has a mean value of about 1.3.  

A caveat to determining a mean value of IR/X for general
application to starbursts arises from the question of whether to
include M\,82.  In Table 1, this source has unusually large values of
IR/X.  We noted in section 2 that the X-ray flux of M\,82 can be
greater by a factor of two when the variable source is in a high
state.  Such a situation would cause the IR/X of M\,82 to be closer to
that of the other starbursts.  Since M\,82 is only one of 9 regions in
Table 1, we decided to include it when defining the mean IR/X for all
starbursts, but we note the slight change that
would occur without M 82. 

The overall mean IR/X, for all redshifts z$>$1 and for all objects, is
1.33.  This reduces to 1.08 without M\,82.  Wide field SIRTF
``shallow'' surveys at 24\,$\mu$m, which can cover about 0.2\,deg$^2$
per hour of observing, should reach sensitivities of 0.4\,mJy.  For
the mean IR/X, a SIRTF starburst source of 0.4\,mJy should have a
typical CXO flux in the observed 0.5--2.0\,keV soft band of
0.3$\times$10$^{-16}$\,ergs\,s$^{-1}$\,cm$^{-2}$. The CXO can reach this flux limit for on-axis individual sources
with $\sim1$~Ms of exposure \citep{Brandt01} and $\sim2$~Ms over a larger field 
\citep{Alexander03}. We can conclude, therefore, that both CXO and SIRTF
should be able to see the same starbursts for z$>$1.  The number of
such objects expected in a field depends significantly on models for
luminosity function and the evolution of ultraluminous infrared galaxies.
The models of Dole et al. (2003) predict about 200 starburst galaxies
deg$^{-2}$ with z$>$1.1, to the 0.4\,mJy limit.  This would correspond
to 25 sources in the 450\,arcmin$^2$ field covered by the CDF-N
survey.

A major motivation for our attempt to determine IR/X for starbursts is to determine whether a comparison of SIRTF and CXO measurements can discriminate between
starburst and AGN as the luminosity source for individual ULIRGs.  This
seems a reasonable expectation, because it is well established that
nearby sources known to be dominated by AGN luminosity systematically have lower
far-infrared to X-ray luminosity ratios than do starbursts
\citep[e.g.][]{Levenson01}.  For type 1 AGN, in which
the broad line region and accretion disk can be observed directly, the
luminosity in hard X-rays (which is what would be seen by CXO in the
observer's frame at z$>$1) is typically 1000 times greater relative to
the far-infrared luminosity than in starbursts.  Even for type 2 AGN,
in which the vicinity of the accretion disk is obscured from direct
view, the analogous factor between AGN and starbursts is about 100,
which implies that some hard X-rays from the AGN penetrate the
obscuring material or escape via scattering from regions above the
obscuring torus.  Sources that are composite, with a type 2 AGN accompanied
by comparably strong circumnuclear star formation, have X-ray to infrared
ratios similar to starbursts.  This is not surprising, since in such
cases much of the X-ray and infrared luminosity arises from the
starburst. While these results derived from considerations of far-infrared luminosity do not necessarily apply for SIRTF observations at 24\,$\mu$m, it would be reasonable to conclude that an AGN is present in any object having IR/X lower than in pure starbursts. 
				
Based on the results and trends in Table 1 and Figure 3, 95\% of the starburst regions considered at the various redshifts have IR/X$>$0.2 (only 2 of the 45 entries in Table 1
are less than 0.2, and all are for the same low luminosity source). It is reasonable, therefore, to consider 
a limiting value of 0.2 as defining the lowest IR/X for a pure starburst.  Any
source with IR/X$<$0.2 can be considered with high probability to contain an AGN which
contributes to the luminosity, because of the additional X-ray power.  For
a SIRTF 0.4\,mJy source, this limiting IR/X means that a starburst
should be fainter than a CXO flux of
2$\times$10$^{-16}$\,ergs\,s$^{-1}$\,cm$^{-2}$.  If the source were brighter than this in the CXO soft band, it can be reasonably
classified as having some luminosity arising from an AGN rather than
being purely a starburst.

This result does not mean, however, that any source with IR/X$>$0.2 can only be powered by a starburst. 
There is a category of obscured AGN in which the intrinsic X-ray
luminosity is totally blocked to $\sim10$~keV by the obscuring
material ($\rm{N_{H}>10^{24}~cm^{-2}}$) and only part of the scattered
component may be seen.  These are the ``Compton Thick AGN''
\citep[e.g.][]{Matt00}.  In this case, IR/X might achieve high values compared to expectations for AGN 
because the absorbed X-ray luminosity would reappear as infrared
luminosity.  \citet{Risaliti99} and \citet{Levenson01} conclude that
known examples of Compton Thick AGN consistently have circumnuclear
starbursts and suggest that these starburst regions could be the
source of the material obscuring the AGN.  If the AGN is a significant
additional contributor to the infrared luminosity but not to the X-ray
luminosity, the IR/X could even be greater than for a pure starburst.  

We do not have an adequate sample of
AGN to compare empirically with the IR/X measured for starbursts.  There are only
three AGN for which
ISO and CXO data are available to analyse in the same way as the
starburst sample; these are NGC\,1068, the Circinus galaxy, and NGC\,6240.\footnote{The
mid-IR data on NGC\,1068 and NGC\,6240 were presented in
\citet{LeFloch01} and \citet{Charmandaris99} respectively, while the
data on Circinus were retrieved from the ISO archive.}  It happens that all three of these are obscured, Compton-thick AGN with circumnuclear
starbursts \citep{Levenson01,Komossa03,Lira02,Sambruna01} so it might
be expected that these would be difficult objects to distinguish from
pure starbursts based only on IR/X.  For these objects, we measured
the smallest feasible spatial region for applying the ISO and CXO spectra
($6''\times6''$) to isolate the AGN as much as possible in hopes of determining IR/X as it applies purely to an AGN component.
Results for these three AGN are presented in Table 3.  NGC\,6240
falls within the AGN criterion already defined (IR/X$<$0.2) for most
redshifts included, the Circinus galaxy is within this criterion at
the highest redshift, but NGC\,1068 would be confused with a starburst
at any redshift based only on IR/X.  This result indicates that IR/X
does not provide an unambiguous criterion for a pure starburst.
Nevertheless, if the infrared luminosity of a composite source
(obscured AGN plus starburst) is dominated by the starburst, it would
be appropriate to include such a source in the census of star formation derived from
infrared surveys.  We conclude, therefore, that the overlap in IR/X
between pure starbursts and NGC 1068 or the Circinus galaxy should not
discourage the use of IR/X as a primary sorting mechanism to
distinguish between starbursts and AGN as the power source of ULIRGs
discovered by SIRTF and Chandra. 

Based on the foregoing analysis, what would be the most interesting and puzzling extragalactic sources to be revealed in comparisons of SIRTF and Chandra surveys?
We have concluded empirically that sources with 3.0$<$IR/X$<$0.2 are starbursts or obscured AGN with circumnuclear starbursts.  Objects with IR/X$<$0.2
have an additional source of X-ray emission beyond a starburst, so can be considered as having a dominant AGN. From
Table 1 and Figure 3, the upper bound for IR/X is about 4, if we ignore the single higher value for M82 for reasons discussed previously.  Objects which would be anomalous according to any of the criteria defined would, therefore, be any objects with IR/X$>$4. If any such objects are found in comparisons of infrared and X-ray surveys, we recommend that they receive particular attention to determine their nature.

\section{Summary}

Archival ISO CVF spectra and archival CXO spectra for several pure
starbursts were evaluated to predict the ratio IR/X of SIRTF
24\,$\mu$m flux densities (in mJy) to CXO total soft band fluxes
(0.5--2.0\,keV flux in units of 10$^{-16}$\,ergs\,s$^{-1}$\,cm$^{-2}$)
which would be observed for spectrally similar starbursts seen at
1$<$z$<$3.  For starbursts, IR/X was determined to have a mean value
of 1.3 and falls between the extremes of 0.2 and 3.0.  These empirical
values for IR/X are suggested as an initial criterion for sorting
ultraluminous infrared galaxies detected in SIRTF and CXO surveys into
sources powered primarily by starbursts compared to those powered
primarily by AGN. Sources with IR/X$<$0.2 probably have a dominant AGN; sources with IR/X$>$4 would be especially interesting, because they have no counterparts in the sample we have examined. 

\acknowledgments

We acknowledge NASA for supporting this work through the Chandra X-ray 
Center (NAS 8-39073) and the SIRTF Infrared Spectrograph team (JPL 960803); AZ also acknowledges support by NASA LTSA grant NAG5-13056 and NASA grant G01-2120X. We thank an anonymous referee for helpful suggestions.

\clearpage

%% Tables should be submitted one per page, so put a \clearpage before

\clearpage

\begin{deluxetable}{lrccccc}
%\rotate
\tabletypesize{\scriptsize} 
\tablecaption{SIRTF flux densities compared to CXO total fluxes for Starbursts\tablenotemark{a}\label{tbl_starburst}}
\tablewidth{0pc}
\startdata \\
\tableline 
\tableline \\
Region & & \multicolumn{5}{c}{Redshift (z)}\\ 
\tableline\\
 & 			& 1		& 1.5		& 2		&2.5		& 3 \\
 & CXO band (keV):	& 1.0--4.0	& 1.25--5.0	& 1.6--6.0	& 1.75--7.0	& 2.0--8.0\\
\tableline \\

Antennae - Knot A  & $f_x$ 		& 530		& 399		& 342		& 315		& 284  \\
	   & $\lambda_{eff}$	& 11.79		& 8.89		& 7.74		& 6.57		& 6.19 \\
	   & $f_{\nu}$		& 62		& 18		& 53		& 16		& 24 \\
	   & IR/X		& 0.234		& 0.113		& 0.465		& 0.178		& 0.338 \\
	     				  		  		  		  
NGC\,4038   & $f_x$ 		& 325		& 271		& 242		& 224		& 206 \\
	   & $\lambda_{eff}$	& 11.48		& 8.87		& 7.77		& 6.57		& 6.16 \\
	   & $f_{\nu}$		& 322		& 111		& 292		& 71		& 93 \\
	   & IR/X		& 1.98		& 1.024		& 3.62		& 1.11		& 1.81 \\
	     				  		  		  		  
NGC\,4039   & $f_x$ 		& 529		& 398		& 341		& 314		& 282 \\
	   & $\lambda_{eff}$	& 11.43		& 9.02		& 7.74		& 6.57		& 6.19 \\
	   & $f_{\nu}$		& 181		& 51		& 165		& 45		& 67 \\
	   & IR/X		& 0.684		& 0.320		& 1.45		& 0.502		& 0.95	 \\	
	     				  		  		  		  
Antennae - Total  & $f_x$		& 7220		& 7030		& 7220		& 7520		& 7760 \\
	   & $\lambda_{eff}$	& 11.43		& 8.89		& 7.77		& 6.63		& 6.22 \\
	   & fv			& 5660		& 1880		& 5250		& 1140		& 1410 \\
	   & IR/X		& 1.57		& 0.670		& 2.18		& 0.529		& 0.730 \\
	     				  		  		  		  
Arp\,220	   & $f_x$		& 441		& 527		& 617		& 704		& 783 \\
	   & $\lambda_{eff}$	& 12.21		& 8.69		& 7.82		& 6.66		& 5.90 \\
	   & $f_{\nu}$		& 351		& 239		& 510		& 116		& 125 \\
	   & IR/X		& 1.59		& 1.13		& 2.48		& 0.577		& 0.639 \\
	     				  		  		  		  
IC\,694	   & $f_x$		& 751		& 855		& 930		& 1000		& 1050 \\
	   & $\lambda_{eff}$	& 12.10		& 8.64		& 7.80		& 6.54		& 6.19 \\
	   & $f_{\nu}$		& 698		& 484		& 968		& 241		& 230 \\
	   & IR/X		& 1.86		& 1.42		& 3.12		& 0.844		& 0.876 \\
	     				  		  		  		  
NGC\,3690   & $f_x$		& 3050		& 3040		& 3110		& 3200		& 3250 \\
	   & $\lambda_{eff}$	& 11.79		& 8.84		& 7.80		& 6.60		& 6.01 \\
	   & $f_{\nu}$		& 1774		& 1091		& 1779		& 605		& 319 \\
	   & IR/X		& 1.16		& 0.897		& 1.72		& 0.662		& 0.393 \\
	     				  		  		  		  
Arp\,299	- Total   & $f_x$		& 6390		& 6000		& 5950		& 6040		& 6040 \\
	   & $\lambda_{eff}$	& 11.89		& 8.79		& 7.80		& 6.57		& 5.99 \\
	   & $f_{\nu}$		& 2238		& 1370		& 2753		& 863		& 429 \\
	   & IR/X		& 0.70		& 0.571		& 1.39		& 0.500		& 0.284	 \\
	     				  		  		  		  
M\,82	   & $f_x$		& 56800		& 56700		& 57400		& 61800		& 61700 \\
	   & $\lambda_{eff}$	& 11.74		& 8.74		& 7.71		& 6.63		& 6.13 \\
	   & $f_{\nu}$		& 68300		& 58400		& 139100	& 28200		& 42200 \\
	   & IR/X		& 2.41		& 2.58		& 7.27		& 1.60		& 2.74 \\

\tablenotetext{a}{ $f_x$ is the X-ray flux of the starburst region in the CXO band shown in units
of 10$^{-16}$ ergs s$^{-1}$ cm$^{-2}$; $\lambda_{eff}$ is the rest
frame wavelength in $\mu$m that would be observed with the MIPS 24\,$\mu$m filter at the redshift listed;
$f_{\nu}$ is the rest-frame flux density of the starburst region in mJy (1\,mJy =
10$^{-26}$\,ergs\,s$^{-1}$\,cm$^{-2}$\,Hz$^{-1}$) at $\lambda_{eff}$; IR/X would be the
observed ratio of SIRTF 24\,$\mu$m flux density in mJy to CXO soft
band (0.5--2.0\,keV) flux in units of 10$^{-16}$ ergs s$^{-1}$
cm$^{-2}$ for a starburst having the spectral shape of this region and sufficient luminosity to be detected at the redshift listed.}

\enddata
\end{deluxetable}

\newpage 

\clearpage

\begin{deluxetable}{lccc}
%\rotate
%\tabletypesize{\scriptsize} 
\tablecaption{Luminosities of Starbursts\label{tbl_irx}}
\tablewidth{0pc}
\startdata \\
\tableline 
\tableline \\
Region		 		& IR/X		& 	L$_x$(2--8\,keV)\\ 
				& (mean in Table 1)	& (ergs\,s$^{-1}$)\\
\tableline \\
Antennae - Knot A		& 0.27			& 2.1$\times 10^{39}$\\
NGC\,4039			& 0.78			& 2.0$\times 10^{39}$\\
NGC\,4038			& 1.91			& 1.5$\times 10^{39}$\\
Antennae - Total		& 1.14			& 5.6$\times 10^{40}$\\
Arp\,220			& 1.28			& 5.4$\times 10^{40}$\\
IC\,694				& 1.62			& 2.2$\times 10^{40}$\\
NGC\,3690			& 0.97			& 6.8$\times 10^{40}$\\
Arp\,299 - Total		& 0.69			& 1.3$\times 10^{41}$\\
M\,82				& 3.32			& 8.0$\times 10^{39}$\\

\enddata
\end{deluxetable}

\newpage
\clearpage

\begin{deluxetable}{lrccccc}
%\rotate
\tabletypesize{\scriptsize} 
\tablecaption{SIRTF flux densities compared to CXO total fluxes for AGN\tablenotemark{a}\label{tbl_agn}}
\tablewidth{0pc}
\startdata \\
\tableline 
\tableline \\
AGN & & \multicolumn{5}{c}{Redshift (z)}\\ 
\tableline\\
 & 			& 1		& 1.5		& 2		&2.5		& 3 \\
 & CXO band (keV):	& 1.0--4.0	& 1.25--5.0	& 1.6--6.0	& 1.75--7.0	& 2.0--80\\
\tableline \\

Circinus   & $f_x$ 		& 15800		& 22600		& 31500		& 73100		& 85000 \\
	   & $\lambda_{eff}$	& 11.79		& 9.04		& 7.69		& 6.68		& 5.86 \\
	   & $f_{\nu}$		& 12450		& 3930		& 8040		& 4360		& 2780 \\
	   & IR/X		& 1.58		& 0.435		& 0.77		& 0.209		& 0.131 \\
	   
NGC\,1068   & $f_x$ 		& 28000		& 21600		& 20500		& 30300		& 31100 \\
	   & $\lambda_{eff}$	& 11.63		& 9.12		& 7.74		& 6.63		& 5.77 \\
	   & $f_{\nu}$		& 33700		& 19000		& 18300		& 13400		& 10900 \\
	   & IR/X		& 2.41		& 2.20		& 2.68		& 1.55		& 1.41 \\	
	   
NGC\,6240   & $f_x$ 		& 4660		& 5590		& 6480		& 10200		& 11000 \\	
	   & $\lambda_{eff}$	& 11.95		& 8.79		& 7.79		& 6.65		& 5.93 \\
	   & $f_{\nu}$		& 454		& 330		& 625		& 198		& 120 \\
	   & IR/X		& 0.20		& 0.15		& 0.29		& 0.068		& 0.044 \\

\tablenotetext{a}{ $f_x$ is the X-ray flux of the AGN in the CXO band shown in units
of 10$^{-16}$ ergs s$^{-1}$ cm$^{-2}$; $\lambda_{eff}$ is the rest
frame wavelength in $\mu$m that would be observed with the MIPS 24\,$\mu$m filter at the redshift listed;
$f_{\nu}$ is the rest-frame flux density of the AGN in mJy (1\,mJy =
10$^{-26}$\,ergs\,s$^{-1}$\,cm$^{-2}$\,Hz$^{-1}$) at $\lambda_{eff}$; IR/X would be the
observed ratio of SIRTF 24\,$\mu$m flux density in mJy to CXO soft
band (0.5--2.0\,keV) flux in units of 10$^{-16}$ ergs s$^{-1}$
cm$^{-2}$ for an AGN having the spectral shape of this AGN and sufficient luminosity to be detected at the redshift listed.}
\enddata
\end{deluxetable}

%
%  Captions for Figures 
% 
 
\clearpage 
\begin{figure} 
\figurenum{1} 
%\plotone{f1.eps}

\caption{ ISO 15\,$\mu$m images of the galaxies in our sample obtained
using the broadband LW3 (12--18$\mu$m) filter of ISOCAM. The
individual areas from where the mid-IR spectra were extracted are
indicated with boxes.  The same areas were used to extract the X-ray
spectra from CXO-ACIS observations. For NGC 4038/39, the large box is Antennae-Total, the upper small black box is NGC 4038, the lower small black box is NGC 4039, and the white box is Antennae-Knot A. For Arp 299, the large box is Arp 299-Total, the medium box is NGC 3690 and the smallest box is IC 694\label{image}.}
\end{figure} 

\clearpage 
\begin{figure} 
\figurenum{2} 
\plotone{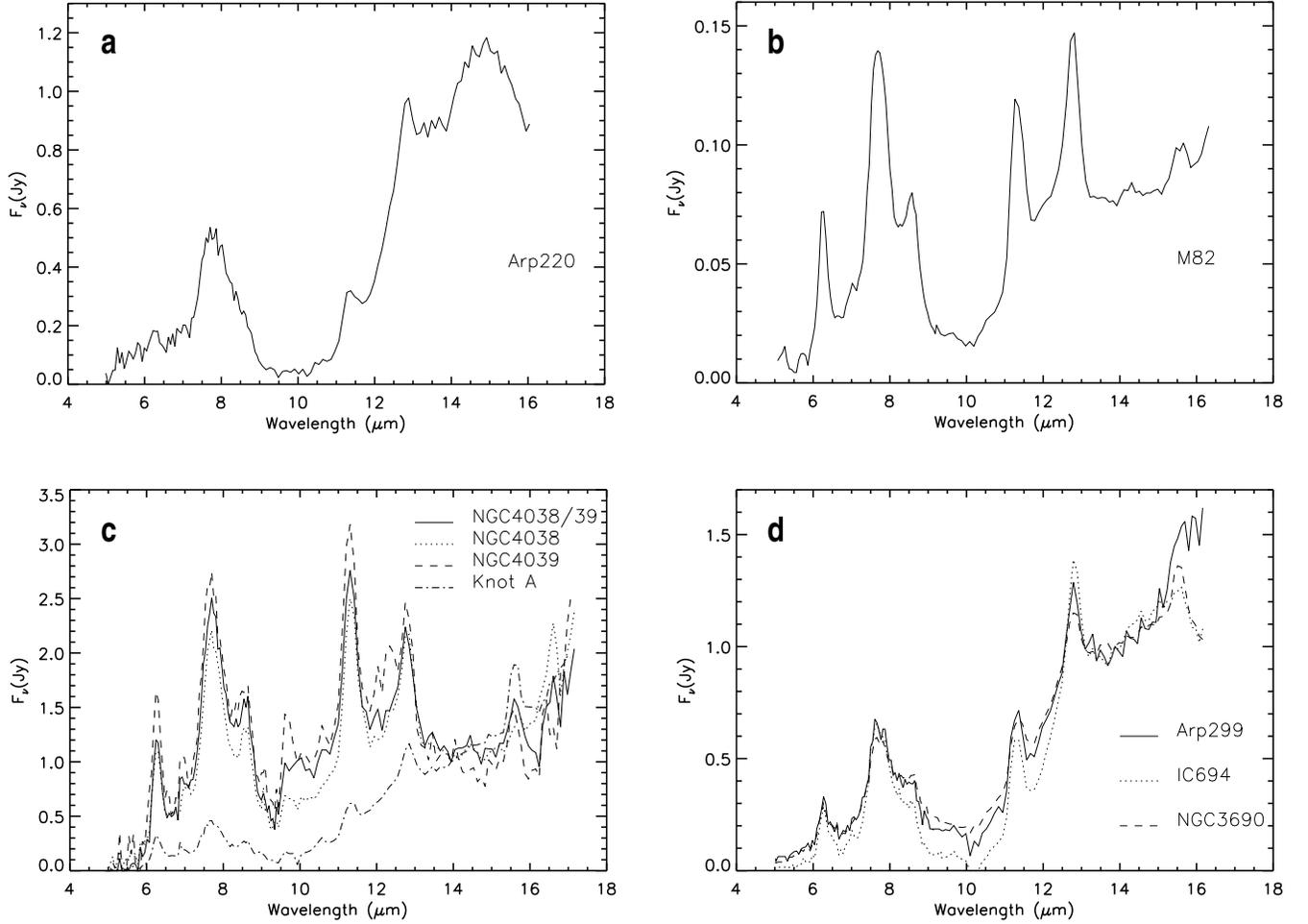}

\caption{
a) Rest frame ISOCAM/CVF 5--16.3\,$\mu$m spectrum of Arp\,220. b)
As in a) but for M\,82. c-d) As in a) but for the various regions
of NGC\,4038/39 and Arp\,299 respectively. Those regions are marked in
Figure 1. Note that in figures 2c) and 2d) the spectra have been
normalized to unity at 14\,$\mu$m. The scale factors by which each
spectrum must be multiplied in order to retrieve the exact flux
density detected are: NGC\,4038=0.138, NGC\,4039=0.060, Knot A=0.116,
NGC\,4038/39=2.153, IC\,694=1.536, NGC\,3690=2.099, and
Arp\,299=4.340.
\label{spectra}}

\end{figure} 

\clearpage 
\begin{figure} 
\figurenum{3} 
\plotone{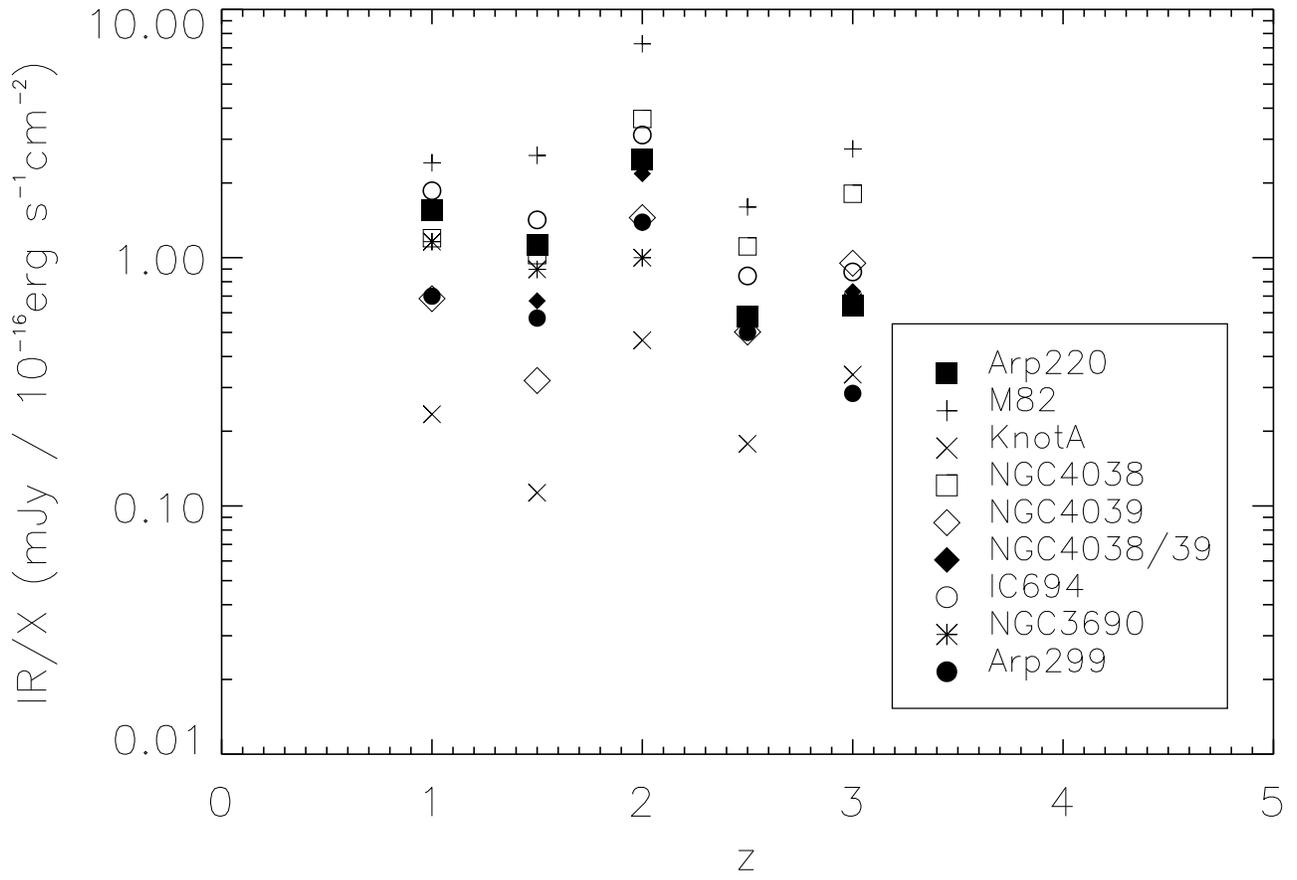}

\caption{The IR/X ratio for the various starburst regions of the sample as a function of redshift.\label{irx}}

\end{figure} 

\end{document}